\input{aipcheck}

\documentclass[
    ,draft            
  ]
  {aipproc}

\layoutstyle{6x9}

\begin{document}

\title{Nonequilibrium structures and dynamic transitions 
in driven vortex lattices with disorder}



\author{Alejandro B. Kolton}
{
  address={Centro At\'{o}mico Bariloche and Instituto Balseiro, 
  8400 San Carlos de Bariloche, Argentina}
}

\author{Daniel Dom\'{\i}nguez}{
  address={Centro At\'{o}mico Bariloche and Instituto Balseiro, 
  8400 San Carlos de Bariloche, Argentina}
}


\begin{abstract}
We review our studies of elastic lattices driven by an 
external force $F$ in the presence of random disorder,
which correspond to the case of vortices in
superconducting thin films driven by external currents. 
Above a critical force $F_c$ we find two  dynamical phase
transitions at $F_p$ and $F_t$, with $F_c<F_p<F_t$.
At $F_p$ there is a transition
from plastic flow to smectic flow
where  the noise is isotropic
 and there is a peak in the differential resistance.
At $F_t$ there is a sharp transition to a frozen transverse solid
where both the transverse noise and the diffussion fall down abruptly
and therefore the vortex motion is localized in the transverse direction. 
From a generalized  fluctuation-dissipation relation 
we calculate an effective
transverse temperature  in the fluid moving phases. 
We find that the effective temperature 
decreases with increasing driving force and
becomes equal to the equilibrium melting temperature  
when the dynamic transverse freezing occurs.
\end{abstract}

\maketitle


\section{Introduction}

The study of the collective motion of vortex lattices in superconductors
has brought new concepts in the non-equilibrium statistical
physics of driven disordered media
\cite{KV,plastic,filam,olson,GLD,BMR,SV,BF,%
bhatta,hellerq,pardo,%
SIM,kolton}.
The prediction by Koshelev and Vinokur
\cite{KV} of a {\it dynamical phase transition}
upon increasing drive, from
a fluid-like flow  regime \cite{plastic,filam,olson} 
to a coherently moving solid \cite{KV}, 
 has motivated an outburst
of theoretical \cite{GLD,BMR,SV,BF}, 
experimental \cite{bhatta,hellerq,pardo}, and simulation
\cite{SIM,kolton} work. 
The relevant physics of the
high velocity driven phase is controlled by the transverse displacements
(in the direction perpendicular to the
driving force) \cite{GLD}, leading to a new class of
driven systems characterized by {\it anisotropic spatial structures} 
with transverse periodicity \cite{GLD,BMR,SV}.
These moving anisotropic vortex structures have been observed 
experimentally by Pardo {\it et al.} \cite{pardo}, and their different
features have been studied  in simulations \cite{SIM,kolton}.

 Koshelev and Vinokur \cite{KV} defined a
``shaking'' temperature $T_{\tt sh}$ from the fluctuating force 
felt by a vortex
configuration moving in a random pinning potential landscape.
This lead to their prediction of a dynamic phase transition
when $T_{\tt sh}$  equals the
equilibrium melting temperature of the vortex system \cite{KV}.
This dynamic transition separates  the
liquid-like phase of vortices moving at low driving forces 
from the ``crystalline'' vortex lattice moving at
large forces. 
However, later work \cite{GLD,BMR,SV} has shown that the perturbation 
theory used in \cite{KV} breaks down and that the vortex phase at large 
velocities can be an anisotropic transverse glass instead of a crystal.
In spite of this, the  shaking temperature introduced in \cite{KV}
has been a useful qualitative concept, at least phenomenologically.

Cugliandolo, Kurchan and Peliti~\cite{CKP} have 
introduced the notion of time-scale dependent 
``effective temperatures'' $T_{\tt eff}$ from a
modification of the fluctuation-dissipation theorem (FDT)
in slowly evolving out of equilibrium systems.
$T_{\tt eff}$  is defined 
from the slope of the parametric plot of the 
integrated response against the correlation function
of a given pair of observables when the latter is bounded or,
equivalently, it is the inverse of 
twice the slope of the parametric plot of the 
integrated response against the displacement
when the correlation is unbounded.
This definition yields a bona fide temperature in the thermodynamic sense 
since it can be measured with a thermometer, it controls the direction 
of heat flow for a given time scale and it satisfies a
zero-th law within each time scale.
$T_{\tt eff}$ was 
found analytically in mean-field glassy models~\cite{CK}
and it was  successfully studied in structural and spin
glasses, both numerically~\cite{teff_sim}
and experimentally \cite{teff_exp}, in granular matter
\cite{teff_gran} and in weakly driven sheared fluids
\cite{shear}.

In this article, we will review our work on the dynamical regimes
and nonequilibrium transitions 
of the driven vortex lattices as a function of the driving force
\cite{kolton},
and show how an adequate definition of
an effective temperature (based on Ref.~\cite{CKP})
can help to understand these phenomena \cite{kolton_teff}.

\section{Model}

The equation of
motion of a vortex in position ${\bf r}_i$ is:
\begin{equation}
\eta \frac{d{\bf r}_i}{dt} = -\sum_{j\not= i}{\bf\nabla}_i U_v(r_{ij})
-\sum_p{\bf \nabla}_i U_p(r_{ip}) + \bf{F}+ 
{\bf \zeta}_i(t)
\end{equation} 
where $r_{ij}=|{\bf r}_i-{\bf r}_j|$ is the distance between vortices $i,j$,
$r_{ip}=|{\bf r}_i-{\bf r}_p|$ is the distance between the vortex $i$ and
a pinning site at ${\bf r}_p$, $\eta=\frac{\Phi_0H_{c2}d}{c^2\rho_n}$ is the
Bardeen-Stephen friction and ${\bf F}=\frac{d\Phi_0}{c}{\bf J}\times{\bf z}$
is the driving force due to an applied current ${\bf J}$.
In two-dimensional superconductor 
 the vortex-vortex interaction potential is logarithmic:
$U_v(r)=-A_v\ln(r/\Lambda)$, with $A_v=\Phi_0^2/8\pi\Lambda$
\cite{filam}.
The vortices interact with a random uniform distribution of
attractive pinning centers with 
$U_p(r)=-A_p e^{-(r/r_p)^2}$ with $r_p$ being the coherence length. 
We normalize length scales by $r_p$, energy scales by $A_v$, 
and time is normalized by 
$\tau=\eta r_p^2/A_v$.   The effect of a thermal bath 
at temperature $T$ is given by the stochastic
force ${\bf\zeta}_i(t)$,  satisfying $\langle {\zeta}^\mu _i(t) 
\rangle=0$ and $\langle {\zeta}^\mu   _i(t){\zeta}^{\mu '}_j(t') 
\rangle = 2 \eta k_B T \delta(t-t') \delta_{ij} \delta_{\mu \mu '}$.
We consider $N_v$ vortices and $N_p$ pinning
centers in a rectangular box of size $L_x\times L_y$, 
and the normalized  vortex density is $n_v=N_vr_p^2/L_xL_y=Br_p^2/\Phi_0$.
Moving vortices induce a total electric field  ${\bf
E}=\frac{B}{c}{\bf v}\times{\bf z}$, with ${\bf v}=\frac{1}{N_v}\sum_i 
{\bf v}_i$.
We use periodic boundary
conditions and the periodic 
long-range logarithmic interaction is evaluated with 
an exact and fast converging sum \cite{log}.
Typical simulation parameters correspond to 
vortex densities of $n_v=0.05-0.12$ in
a box with $L_x/L_y=\sqrt{3}/2$, with 
$N_v=64-784$, 
pinning strengths of $A_p/A_v=0.1-0.35$, and
densities of pinning centers $n_p > n_v$ 
(dense pinning $n_p>n_v$, is 
typically realized in experimental samples). 
The equations are integrated
with a time step of $\Delta t=0.01-0.1\tau$ and averages are
evaluated in $30000 - 80000$ integration steps after $2000-30000$ 
iterations for  equilibration (when the total energy reaches a 
stationary value). 


\section{Dynamical Regimes}

We start  by showing the vortex trajectories and their 
translational order in the different steady state regimes.
 In the upper panel of Fig.~1
 we show the vortex trajectories $\{ {\bf r}_i(t)\}$ 
for typical
values of $F$ by plotting all the positions of the vortices for all the
time iteration steps. 
We also study the 
time-averaged structure factor 
$S({\bf k})= \langle|\frac{1}{N_v}\sum_i \exp[i{\bf k}\cdot{\bf
r}_i(t)]|^2\rangle$, which is shown in the middle pannel of Fig.~1.
The number of deffects in the lattice structure can be obtained
from a Voronoi construction. This in shown in the lower pannel of
Fig.~1, where the lattice defects are shown in gray.
In Fig. 2(a) we plot  the average vortex velocity 
$V=\langle V_y(t)\rangle=\langle\frac{1}{N_v}\sum_i \frac{dy_i}{dt}\rangle$,
 in the 
direction of the force as a function of $F$ and its corresponding
derivative $dV/dF$.  
Below a critical force $F_c$ all the vortices are
pinned and 
there is no motion, $V=0$. Above $F_c$, we can distinguish
three different dynamical regimes:

\begin{figure}
  \includegraphics[height=.45\textheight,angle=90]{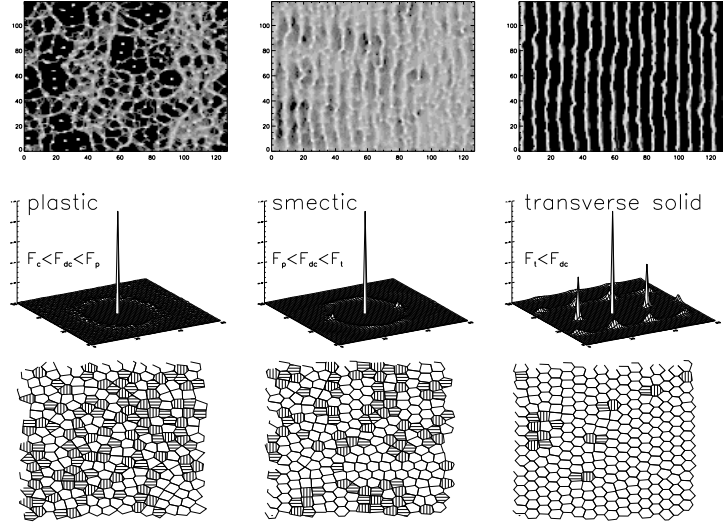}
  \caption{Examples of steady states for different dynamical regimes.
  Upper panel: Vortex trajectories. Middle pannel: Average structure
  factor $S({\bf k})$. Lower pannel: Voronoi construction for each given
  vortex structure.}
\end{figure}

{\it (i) Plastic flow}: $F_c<F<F_p$.
At $F_c$ vortices start to move in 
a few filamentary channels  \cite{filam}.
A typical situation is shown in Fig. 1, where 
a fraction of the
vortices are moving in an intricate network of channels. 
As the force is increased a higher fraction of  vortices
is moving.
In this regime, vortices can move in the transverse direction
(perpendicular to ${\bf F}$) through the tortuous structure
of channels \cite{olson}.
We see that the corresponding $S({\bf k})$ only has the central peak showing
the absence of ordering in this plastic flow regime \cite{plastic,filam,olson}
and that there are a large number of defects.

{\it (ii) Smectic flow}: $F_p<F<F_t$.
We find a peak in the
differential resistance, $dV/dF$, at a characteristic force $F_p$.
At $F=F_p$ we see that
{\it all} the vortices are moving in a 
seemingly isotropic channel network 
with maximum interconnectivity.
In the experiment of Hellerqvist {\it et al} \cite{hellerq}
the value of $F_p$ was taken as an indication of
a dynamical phase transition.
In fact, we 
find that above $F_p$ a new dynamic regime sets in. In this 
case, as we show in Fig. 1, all the vortices are moving
in trajectories that are mostly parallel to the force, 
forming ``elastic channels''. 
Two small Bragg peaks appear in the
structure factor along the $K_y=0$ axis,
which correspond to ${\bf G}_1=(\pm 2\pi/a_0,0)$. 
This is consistent with the onset of ``smectic'' ordering
\cite{BMR}
 in the transverse direction with elastic channels separated by 
 a distance  $\sim a_0 = n_v^{-1/2}$.
In this regime the transverse motion consists in
vortex jumps from one channel to another, resembling
``thermally'' activated transitions induced by local chaos. 
The rate of these transitions
decreases with increasing force. 
In Fig. 2(b) we plot the magnitude 
of the Bragg peaks at $G_1$, $S_s=S(G_1)$, corresponding to
smectic ordering ($K_y=0$), and the other neighboring peaks
at ${\bf G}_2=\pm2\pi/a_0(1/2,\sqrt{3}/2)$ and
${\bf G}_3=\pm2\pi/a_0(-1/2,\sqrt{3}/2)$
, $S_{l}=(S(G_2)+S(G_3))/2)$, corresponding
to longitudinal ordering ($K_y\not=0$).
We see that above $F_p$ the 
intensity of the smectic peak $S_s$  starts to grow and $S_s\gg S_l$,
while below $F_p$ the spatial structure is isotropic, $S_s=S_l\ll 1$.
The Bragg peak heights depend with system size 
as $S(G) \sim N_v^{-\sigma_G}$, where
$\sigma_G=0$ means long-range order (LRO), $0<\sigma_G<1$ means
quasi long-range order (QLRO) and $\sigma_G\ge 1$ means short-range order
(SRO). We have found that  $\sigma_G \ge 1$ in this regime: there is only
short range smectic order \cite{kolton},
and thus this phase corresponds to a fluid. 
In this sense the
transition at $F_p$ is a dynamic transition in the flow.

{\it (iii) Frozen transverse solid}: $F>F_t$.
At a new characteristic force $F_t$, the 
jumps between channels suddenly stop and vortex motion becomes
frozen in the direction perpendicular to ${\bf F}$. An example for
$F>F_t$ is shown in Fig. 1 where we see well defined elastic channels 
parallel to ${\bf F}$. 
We see that 
new peaks appear in $S({\bf k})$ in directions
with $K_y\not=0$, like $G_2$, $G_3$, showing  that there is some
 longitudinal ordering between the channels.  These
 extra peaks are smaller than the smectic peaks, and $S({\bf k})$ is 
 very anisotropic. 
We note in Fig. 2(a) that $F_t$ is the point where the
noisy behavior in $dV/dF$ ceases.
A similar criterion was used by Bhattacharya and Higgins to define
their dynamical phase diagram \cite{bhatta}.
In Fig. 2(b) we see that in $F_t$
there is an increase in the longitudinal ordering
$S_l$, and both $S_s$ and $S_l$ tend to saturate at an almost constant value
for $F\gg F_t$. Furthermore, we have
we found that there is  QLRO 
with a value of $0.5<\sigma_G <0.7$ \cite{kolton}.
We also see that in this phase there are very few defects,
corresponding to dislocations oriented in the direction of the force. 

\begin{figure}
  \includegraphics[height=.5\textheight]{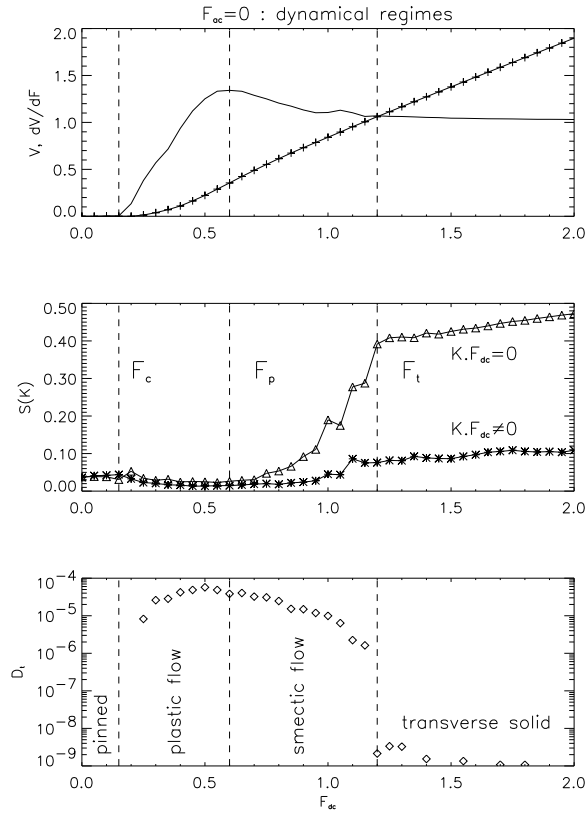}
  \caption{(a) Velocity-force curve (voltage-current characteristics),
$+$ symbols. $dV/dF$ curve (differential resistance),
continuous line.
(b) Intensity of the Bragg peaks.
For smectic ordering,  $K_y=0$ : triangles;
for longitudinal ordering, $K_y\not=0$, asterisks. 
(c) Diffusion coefficient for transversal motion, $D_x$.}
\end{figure}

\section{Dynamical transitions}

 An understanding of the {\it dynamical} transitions
can be obtained from studying the temporal behavior of the system. 
It has been observed experimentally
 that the longitudinal voltage can show low frequency noise.
  This voltage noise reaches a very large
 value above the critical current, which
 has been attributed to plastic flow, and then
 the  noise decreases for large current.
In addition, even when the total dc transverse
voltage $\langle V_x\rangle=
\langle\frac{1}{N_v}\sum_i \frac{dx_i}{dt}\rangle$ is zero, 
it can also have fluctuations and noise. 
In fact, it is easy to understand that this transverse noise
 will be closely related to the wandering and wiggling of
the plastic flow channels and to the jumps between elastic channels
in the smectic phase.
We have calculated in \cite{kolton} the
power spectrum of both the longitudinal voltage
and the transverse voltage.
We have found \cite{kolton} that near the critical force, the 
longitudinal noise, $P_y$, is larger than 
the transverse noise, $P_x$.   
At $F_p$ the voltage noise becomes isotropic, $P_x=P_y$. 
This is the point where we have seen
the highest interconnection in the channel network.
The coincidence of isotropic noise, onset of short-range smectic
order and peak in differential resistance 
suggest that there is a dynamic transition  at $F_p$. 
Above $F_p$, the onset of elastic channels and smectic ordering reduces the
longitudinal noise, while the transverse noise remains large 
due to the ``activated'' jumps between elastic channels. 
At $F_t$ the transverse noise falls abruptly, by several orders of magnitude.
This corresponds to a {\it freezing transition} 
of vortex motion in the transverse
direction. Above $F_t$ there are no more vortex jumps between elastic
channels. The low frequency noise can be closely related to diffusive
motion for large times. We have analyzed the average 
quadratic displacements of vortices in the transverse direction
from their center of mass 
position $(X_{cm}(t),Y_{cm}(t))$ as a function of time. We  define 
$w_x(t)=\frac{1}{N_v}\sum_i[\tilde{x}_i(t)-\tilde{x}_i(0)]^2$,
where $\tilde{x}_i(t)=x_i(t)-X_{cm}(t)$.
We have found that the vortex motion in the fluid 
regimes, for $F_c < F < F_t$, is diffusive
in the transverse direction, with  $w_x(t)\sim D_x t$.
In Fig. 2(c) we show the behavior of the
transverse diffusion coefficient $D_x$.  
The transverse
diffusion is maximum at the peak in the differential resistance, $F_p$,
and it has a clear and abrupt jump to zero at $F_t$ indicating the transverse 
freezing transition. It is interesting to note that melting transitions
also show a jump in the diffusion coefficient.

\section{Effective temperature}

To study the fluctuation-dissipation relation (FDR), 
we proceed in a similar way as in simulations of 
structural glasses \cite{teff_sim} and consider the observables:
\begin{equation}
A_\mu   (t)=\frac{1}{N_v}\sum_{i=1}^{N_v} s_i  r^\mu   _i(t)\;;\;\;\;\;
B_\mu   (t)=\sum_{i=1}^{N_v} s_i  r^\mu   _i(t) \; ,
\end{equation}
where $s_i=-1, 1$ are random numbers with 
$\overline{s_i}=0$ and $\overline{s_i s_j}=\delta_{ij}$, 
and $r^\mu   _i=R^\mu   _i-R^\mu   _{cm}$
with $\mu   =x,y$ and
${\bf R}_{cm}$  being the center of mass coordinate. 
Taking ${\bf F}=F{\bf y}$ 
we  study separately the FDR in the transverse and 
parallel directions with respect to ${\bf F}$. 
The time correlation function between the observables 
$A_\mu   $ and $B_\mu   $ is 
\begin{equation}
C_\mu   (t,t_0) = \overline{\langle A_\mu   (t)B_\mu   (t_0)\rangle} 
=  \frac{1}{N_v}
\sum_{i=1}^{N_v}\langle 
r^\mu   _i(t) r^\mu   _i(t_0)\rangle \; ,
\end{equation}
since the $r^\mu _i$ are independent of the $s_i$. The integrated response 
function $\chi_\mu   $ for the observable $A_\mu   $ is obtained by applying 
a perturbative force ${\bf f}^\mu   _i = \epsilon s_i \hat{\mu   }$ 
(where $\hat{\mu   }=\hat{x}, \hat{y}$)
at time $t_0$ and keeping it constant for all subsequent times
on each vortex:
\begin{eqnarray}
\chi_\mu   (t,t_0) = \lim_{\epsilon \to 0}\frac{1}{\epsilon}
\Bigl[ \overline{\langle {A_\mu   (t)} \rangle}_{\epsilon} - 
\overline{\langle {A_\mu   (t)} \rangle}_{\epsilon=0}
\Bigr] \; .
\end{eqnarray}
We then define a function, $T^\mu   _{\tt eff}(t,t_0)$, by the relation,  
\begin{eqnarray}
\chi_\mu   (t,t_0)=\frac{1}{2k_B T^\mu   _{\tt eff}(t,t_0)} \Delta_\mu 
(t,t_0) \; ,
\end{eqnarray}
where
$\Delta_\mu   (t,t_0) = \frac{1}{N_v}\sum_{i=1}^{N_v}
\langle|r^\mu   _i(t)-r^\mu   _i(t_0)|^2\rangle 
= C_\mu   (t,t)+C_\mu   (t_0,t_0)-2C_\mu   (t,t_0)$
is the quadratic mean displacement 
in the direction of $\hat{\mu   }$. 
For a system in equilibrium at temperature $T$ the FDT
requires that $T^x_{\tt eff}=T^y_{\tt eff}=T$. 
In a nonequilibrium system, like the driven 
vortex lattice with pinning, the FDT does not apply. 
Since we are interested in the {\it stationary} states reached 
by the driven vortex lattice, where aging effects are stopped
\cite{shear},
then all observables  depend only on the difference
$t-t_0$, if  
we choose $t_0$ long enough to ensure stationarity. 
From the parametric plot of $\chi_\mu   (t)$ vs.
$\Delta_\mu   (t)$ one defines the effective temperature 
$T^\mu   _{\tt eff}(t)$ using Eq.~(5),
provided $T^\mu   _{\tt eff}(t)$ is a constant
in each time-scale  \cite{CKP}.

We have studied \cite{kolton_teff} the transverse and longitudinal FDR for the 
moving vortex lattice as a function of driving force, $F$, 
solving the dynamic equations 
for different values of $A_p$, $n_v$, and $T$ within the
fluid regimes, $F_c<F<F_t$. 
We have obtained \cite{kolton_teff} for each $F$ parametric plots of the
transverse 
quadratic mean displacements, $\Delta_x(t)$ as a function
of the
integrated transverse response, 
$\chi_x(t)$.
We found that the equilibrium FDT does not apply in general
but two approximate 
linear relations exist for $\Delta_x(t)< 0.05 r_p^2$ and 
for $ \Delta_x(t) > r_p^2$, with a non-linear crossover between them. 
We have found
that the short displacements region corresponds to the bath temperature, 
$T$, meaning that the equilibrium FDT 
applies in the transverse direction 
only for short times,  $t\ll   r_p/v$. 
For the large displacements region we have obtained  a different
slope corresponding to an effective transverse 
temperature $T_{\tt eff}^x(T) > T$. Furthermore, when  
comparing the results for different $T$, we have found  
$T_{\tt eff}^x(T) \approx T_{\tt eff}^x(0)+T$. 
On the other hand, a similar analysis of
the FDR for the longitudinal direction 
does not show a constant slope
for large displacements or long times such that $t>r_p/v$. 
\begin{figure}
\centerline{
\includegraphics[height=.3\textheight]{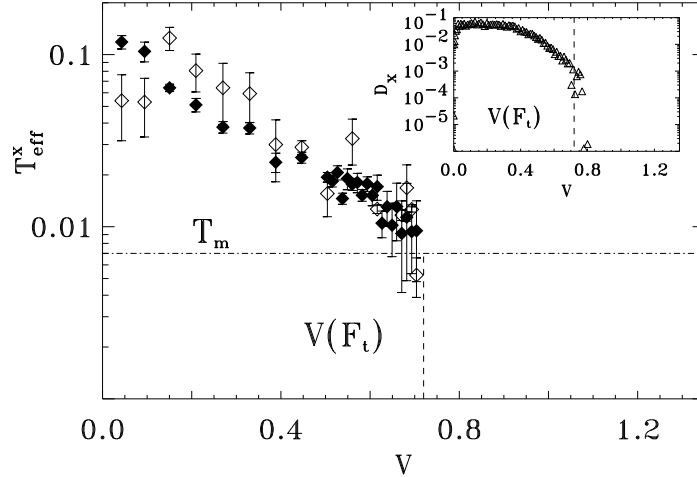}}
\caption{Transverse effective temperature $T_{\tt eff}^x$ vs voltage $V$ 
for $A_p=0.35$, $T=0$ and $n_v=0.07$, using the diffusion relation 
(filled diamonds) and a generalized Kubo formula (open diamonds). Inset: 
Transverse diffusion constant $D_x$ vs $V$. Dashed lines indicate the 
transverse freezing transition at $F=F_t$ and the dash-dotted line 
indicate the melting temperature 
of the unpinned system $T_m \approx 0.007$.}
\end{figure}\noindent

In Fig.~3 we show the transverse effective temperature obtained
in Ref.~\cite{kolton_teff}, 
$T_{\tt eff}^x$, for $T=0$
as a function of voltage ({\it i.e.}, average velocity, $V$). 
We observe that  $T_{\tt eff}^x$ is a 
decreasing function of $V$. The most remarkable result
is  that $T_{\tt eff}^x$ reaches a value close to 
the equilibrium melting 
temperature of the unpinned system, $T_m \approx 0.007$ \cite{tmelting}, when 
the system approaches the transverse freezing transition at $F=F_t$
(obtained from the vanishing of the transverse diffusion 
$D_x$, shown in the inset).  We have found that  
$T_{\tt eff}^x \rightarrow T_m$ when $F \rightarrow F_t$
for different values of 
pinning amplitude, $A_p$, and vortex density, $n_v$,
even when $F_t$ depends on $A_p, n_v$.

\section{Conclusions}

In conclusion, we have obtained evidence
of two dynamical phase transitions.
The first transition at $F_p$ is the point of isotropic
noise and maximum transverse diffusion 
and corresponds to the observed peak in the differential resistance.
The second transition at $F_t$ is a freezing transition in the {\it transverse}
direction, where 
the transverse diffusion vanishes
abruptly.
We have been able to define an effective temperature in
the moving fluid phase, using the thermodynamically adequate
definition of \cite{CKP}.
The fact that we find that
dynamic freezing occurs when $T^x_{\tt eff}(T)=T_m$,
clearly indicates that there is a nonequilibrium phase transition at $F_t$.


\begin{theacknowledgments}

The research work reviewed here is the result of collaborations
with N. Gronbech-Jensen, L. Cugliandolo and R. Exartier.
Our work in Argentina has been supported by 
ANPCYT (grants PICT97-03-00121-02151, PICT97-03-00000-01034, PICT99-03-06343),  by 
Fundaci\'{o}n Antorchas (grant A-13532/1-96),  Conicet (grant PIP-4946/96)  
and CNEA (within program P-5).
A.B.K is also supported  by a fellowship from Conicet.

\end{theacknowledgments}


\bibliographystyle{aipproc}   



\end{document}